\documentclass[acmlarge]{acmart}
\usepackage{booktabs} 
\usepackage{epigraph}
\usepackage{xcolor}
\usepackage{graphicx,calc}
\usepackage{subcaption}
\usepackage[ruled]{algorithm2e} 
\usepackage{csquotes}
\usepackage{wrapfig}

\usepackage{color, colortbl}
\definecolor{Gray}{gray}{0.9}

\colorlet{shades}{red!15!green!10!blue!15}
\colorlet{links}{red!40!green!200!blue!20}
\newcommand{\shade}{\colorbox{shades}} 
 
\newlength\myheight
\newlength\mydepth
\settototalheight\myheight{Xygp}
\begin{document}
 
\title{Blockchain: A Graph Primer} 
\author{Cuneyt Gurcan Akcora}
 
\orcid{***-5678-9012-3456}
\affiliation{%
  \institution{University of Manitoba}
  \streetaddress{65 Chancellors Circle}
  \city{Winnipeg}
  \state{Manitoba}
  \postcode{R3T2N2}
  \country{Canada}}
  \author{Yulia R. Gel}
\orcid{***-5678-9012-3456}
\affiliation{%
  \institution{University of Texas at Dallas}
  \streetaddress{W. Campbell Rd}
  \city{Richardson}
  \state{TX}
  \postcode{75252}
  \country{USA}}
\author{Murat Kantarcioglu}
\orcid{1234-5678-9012-3456}
\affiliation{%
  \institution{University of Texas at Dallas}
  \streetaddress{W. Campbell Rd}
  \city{Richardson}
  \state{TX}
  \postcode{75252}
  \country{USA}}

\begin{abstract}
\textbf{\large{Version 2022:}} Since we submitted it to Arxiv in August 2017, this primer has been cited widely and served as a useful resource for anyone looking to start working on blockchain transaction networks. At the time of submission, academic research on blockchains was in its early stages, with few articles on transaction network analysis available. However, the field has grown significantly in the past few years, with multiple conferences, workshops, and journals dedicated to disseminating blockchain research. Access to blockchain data has also been facilitated by tools such as Bitcoin-ETL by Google and repositories such as \url{https://github.com/Chartalist}.

We are pleased to see that our blockchain predictions have come true. For instance, Ethereum has become a major blockchain, and its asset networks have spawned a billion-dollar industry in Decentralized Finance. DAOs have also emerged as a major area of development. Blockchain research is now a significant and influential field where futures are made and careers are built (including our own). With this updated version, we reaffirm our commitment to the field and hope to help even more researchers. 

\noindent \textbf{\large{Changes:}} We have updated the text to reflect recent developments in blockchains. We have expanded on Bitcoin address types and removed content on P2P network aspects. We have described Decentralized Finance and privacy coins in detail. We have also added information on second-layer solutions and Proof-of-X. Outdated information, such as transaction malleability attacks, has been removed. The original 2017 version can be accessed at \url{https://github.com/cakcora/7570-blockchain/blob/master/files/primer2017.pdf}.

\rule{15cm}{0.4pt}

\noindent\large{\textbf{Abstract}}

Bitcoin and its underlying technology, blockchain, have gained significant popularity in recent years. Satoshi Nakamoto designed Bitcoin to enable a secure, distributed platform without the need for central authorities, and blockchain has been hailed as a paradigm that will be as impactful as Big Data, Cloud Computing, and Machine Learning.

Blockchain incorporates innovative ideas from various fields, such as public-key encryption and distributed systems. As a result, readers often encounter resources that explain Blockchain technology from a single perspective, leaving them with more questions than answers.

In this primer, we aim to provide a comprehensive view of blockchain. We will begin with a brief history and introduce the building blocks of the blockchain. As graph mining is a major area of blockchain analysis, we will delve into the graph-theoretical aspects of Blockchain technology. We will also discuss the future of blockchain and explain how extensions such as smart contracts and decentralized autonomous organizations will function.

Our goal is to provide a concise but complete description of blockchain technology that is accessible to readers with no prior expertise in the field.
\end{abstract}
 
 
\maketitle

\section{Blockchain}
Blockchain is a decentralized and secure database technology that was first proposed by the mysterious author, Satoshi Nakamoto, in 2008. It is composed of blocks of transactions that can be verified and confirmed without the need for a central authority. The technology gained widespread popularity through its use in the digital currency, Bitcoin, where each transaction is financial in nature.

While blockchain originated with Bitcoin, it has since found applications in many other areas. Companies have created blockchain-based products for a range of purposes, from tracking diamonds to distributing food products around the world. This growing interest has been further fueled by another popular blockchain application, Ethereum.

While some aspects of new blockchain applications may differ from Bitcoin (such as asset transactions), these differences are minor. Bitcoin transactions offer useful examples for understanding how blockchain works. Similarly, companies that track and analyze financial transactions on the Bitcoin network often refer to their research efforts as "blockchain analysis." The long history of Bitcoin usage since 2009 has also demonstrated the utility and limitations of blockchain in real-world applications.

Blockchain is well-suited to graph analysis, as public addresses are linked with transactions and transferred amounts can be represented as weighted edges. The core principles behind blockchain also shape the creation and maintenance of nodes and edges in graphs. In addition to the rules imposed by the blockchain protocol, certain user practices can also affect how blockchain graphs are updated over time. For example, it is common for users to move their remaining coins to a new address after each transaction (creating a change address).

In this primer, we aim to provide a concise but complete overview of blockchain for researchers working in graph mining. We will cover the core principles and user practices of blockchain and explain how they impact research efforts.
 
We will start with a brief history of Blockchain in the next section.

\section{A Brief History}
\label{sec:history}

\epigraph{Nakamoto may have been the mother of Bitcoin, but it is a child of many fathers: David Chaum's blinded coins and the fateful compromise with DNB, e-gold's anonymous accounts and the post-9/11 realpolitik, the cypherpunks and their libertarian ideals, the banks and their industrial control policies, these were the whole cloth out of which Nakamoto cut the invention.}{Ian Grigg \cite{griffith2014quick}}

Bitcoin \footnote{In community practice, bitcoin is used for currency units, whereas Bitcoin refers to the software, protocol, and community. } represents the culmination of efforts from many people and organizations to create an online digital currency.  Notable currencies from early times are \textit{ecash} from Chaum and \textit{b-money} from Dai. Up to the seminal Bitcoin paper by Nakamoto \cite{nakamoto2008bitcoin}, multiple times digital currencies seemed to have finally succeeded in creating a viable payment method \cite{griffith2014quick}. Hindered by laws and regulations but mostly due to technical shortcomings, none of these currencies took root. Each failure, however, allowed digital enthusiasts to learn from the experience and propose a new building block towards a viable currency. Bitcoin's popularity brought a flood of alternative digital currencies. The earliest being Litecoin, these {alt-coins} propose modifications to Bitcoin in aspects like block creation frequency and block mining. More fundamental changes have come from products not related to digital currencies. Ethereum, specifically, was designed to be the \textquote{World Computer} to de-centralize and democratize the Internet, a network of thousands of linked computers. We will discuss these improvements in Section \ref{sec:future}.
 
From the beginning, digital currencies ran into several fundamental problems. A significant hurdle was the need for a central authority to keep track of digital payments among users. Companies that invented digital currencies proposed to be the central authorities themselves. In a sense, rather than eliminating financial institutions, currencies tried to replace them.  This approach never gained traction.

An alternative to the central authority approach was to use a distributed, public ledger to track user balances; every user stores account balances of every other user. Although theoretically possible, this solution is unfeasible in practice because information about transactions cannot be digitally transmitted in fast and reliable ways. After all, the networks are faulty and malicious users benefit from lying about balances in double-spending attacks. The problems with achieving consensus in a distributed system, known as the Byzantine general's problem,  has been a well-studied topic in distributed systems~\cite{lamport1982byzantine}.

While the issue of central authority was still a challenge in the digital currency field, a solution was found in the unrelated domain of spam detection. Email providers were facing the problem of receiving a large number of spam emails, which could be analyzed and marked as spam, but this process wasted system resources. Researchers sought a way to check emails for spam without using excessive resources.

Dwork et al. proposed a solution called Proof-of-Work (POW) that involved email senders performing a time-consuming computation and appending proof of the computation to the email. A naive POW would be counting the number of words in the email and attaching the proof (e.g., this email has 35 words) to the end of the email. In this naive POW scheme, an email provider (e.g., Gmail) uses the following algorithm on the incoming emails:  

\begin{enumerate}
    \item If the email does not contain the POW, discard it. The email may be non-spam, but the email provider will not attempt to check its content. 
    \item If the email contains the POW, re-compute the POW to validate the work. If the POW is not valid, discard the email.
    \item If the POW is correct, the email can still be spam. Run the email through a spam detection algorithm and discard it if it is labeled as spam.
    \item If the email has passed both POW and spam filters, deliver the email to its destination account.

\end{enumerate}

The naive word counting method described here is not a good POW system, because it does not require a significant amount of computational effort. A good POW should be difficult to compute, but easy to verify. 

\noindent\textbf{HashCash or Dwork?} Appearing in 1997, HashCash proposed POW to prevent service denial attacks and spam emails~\citep{back2002hashcash}. HashCash authors claim that at the time of their writing, they were unaware of \citep{dwork1992pricing} who also proposed a POW scheme. Furthermore, the Bitcoin white paper cites the HashCash white paper~\citep{back2002hashcash}, cementing the importance of HashCash in POW.

It is important to note that while POW does not prevent all spam emails, it makes it costly to send a large number of them. This helps to reduce the amount of spam email that is sent and received, making the process of checking emails for spam more efficient. Nakamoto used POW to create an efficient block mining process in Bitcoin. Before we explain Bitcoin POW, we need to teach three data structures that Bitcoin uses:

\begin{itemize}
    \item A Bitcoin \textbf{address} is a unique identifier for an account and is typically a short text string. In the case of Bitcoin, addresses are used to identify accounts and facilitate transactions between them. Each address is unique, allowing users to easily send and receive coins without confusion.
\item A Bitcoin \textbf{transaction} is a record of a coin transfer between addresses on the blockchain. In the case of Bitcoin, transactions are used to move coins from one address to another, allowing users to send and receive payments. 
\item A Bitcoin \textbf{block} stores error-free and non-conflicting transactions.
\end{itemize}

To join a blockchain network, a user must install a software application called a wallet and create connections with other network participants to learn about past transactions and blocks. This process does not require user identification, and the wallet software functions similarly to a torrent application, such as BitTorrent. Blockchain network peers include ordinary nodes and miners who attempt to create clean and non-conflicting blocks through a process called block mining. In a distributed setting, such as the one used by Bitcoin, clean blocks are those that do not contain conflicting transactions between valid addresses.

Conflicts can arise due to transaction details and address balances. For example, if a user attempts to spend the same bitcoins in two different transactions, the transactions will conflict with each other. This is known as a double-spending transaction. A miner will choose to include only one of the conflicting transactions in a block, while discarding the other. Double-spending transactions typically imply attempts to defraud Bitcoin users. By ensuring that only clean and non-conflicting transactions are included in blocks, miners help to maintain the integrity and security of the blockchain network.

Bitcoin does not restrict anyone from joining the network, and every peer on the network can attempt to mine a block. However, this freedom could result in too many blocks being created, which would be inefficient. To prevent this, Bitcoin uses a POW mechanism that makes it difficult for miners to create new blocks. This allows miners to create at most one block approximately every 10 minutes. 

There are two essential aspects of POW. First, POW makes it difficult for miners to lie about the contents of blocks; a miner must spend significant computational power to mine a block. Second, the time it takes to create a new block allows network users to learn about the latest blocks and their transactions, which helps to prevent double-spending attacks. In the academic jargon this is called reaching a \textquote{consensus about bitcoin user balances}. In other words, every user knows how many bitcoins a user owns. In addition, each block contains information about the previous block, creating an immutable chain of blocks known as the blockchain. This ensures the security and integrity of the network by linking the latest block to all the previous blocks. Any changes to past blocks can be easily detected, making it difficult for attackers to tamper with the blockchain. 

Before we detail how Bitcoin and its POW work, we need to teach the three fundamental data structures.

\section{Building Blocks of Blockchain}
\label{sec:buildingblocks}

\subsection{Address}
In essence, a blockchain address is a unique string of 26-35 characters that is derived from a public key, which is in turn generated from a private key. Owning an address means having the private keys associated with that address. The private keys are used to sign transactions, proving that the owner of the address is the one initiating the transaction. This ensures the security and authenticity of transactions on the blockchain.

Currently, Bitcoin uses four kinds of addresses:
\begin{itemize}
\item P2PKH (Since 2009): \textit{Pay to PubkeyHash} address that starts with 1 as in {\tiny \shade{1Pudc88gyFynBVZccRJeYyEV7ZnjfXnfKn}}.
\item P2SH (Since 2012): \textit{Pay to ScriptHash} address that starts with 3 as in {\tiny \shade{3J4kn4QoYDj95S3fqajUzonFhLyjjfKjP3}}.
\item Bech32 (Since 2017): \textit{Segregated Witness address} that starts with bc1q as in \\ {\tiny\shade{bc1qc7slrfxkknqcq2jevvvkdgvrt8080852dfjewde450xdlk4ugp7szw5tk9}}.
\item Taproot (Since 2021): \textit{Pay to Taproot (P2TR)} addresses that start with bc1p as in 
{\tiny\shade{bc1pmzfrwwndsqmk5yh69yjr5lfgfg4ev8c0tsc06e}}.
\end{itemize}

The most common (we may even say the standard) type is the \textit{Pay to PubkeyHash}, where we use a single private key to spend bitcoins received from a transaction. Bitcoin added the \textit{Pay to ScriptHash} functionality later to support m-of-n multi-signature transactions; at the receiving address, bitcoins can be spent only when at least m out of n users sign the transaction. In theory, each of n private keys belongs to different users who must come together to spend the coins. In practice, all keys belong to the same user, and the scheme is used to prevent coin theft through stolen private keys. Bech32 is used for Segregated Witness~\footnote{Segregated Witness leaves transaction signatures out of blocks to save space.} outputs. Taproot was designed to make it more difficult to determine whether an address requires multiple signatures to unlock the bitcoins it receives. Despite its potential benefits, Taproot addresses have not yet gained widespread adoption among Bitcoin users. P2SH and P2PKH are case sensitive and Base58 coded, whereas Bech32 is case insensitive and Base32 coded. 

We cannot initiate a transaction without a receiver address, and an error checking code appended to the address prevents address typo errors. It is not possible to send bitcoins to a malformed address. A user may publish its address on web forums, mailing lists, or other mediums. 

Each type of Bitcoin address is created using a specific protocol, but from a technical perspective, the type of address used in a transaction does not matter, as all types of addresses can be used as input or output addresses. However, the use of Pay to ScriptHash addresses in transactions makes it possible for multiple users to participate in the spending decision. This user behavior can be mined for various applications, such as fraud prevention and marketing.

\subsection{Output and Transaction}
In Bitcoin, a transaction is a transfer of coins between addresses. This transfer is represented by sending and receiving a data structure called an \textbf{output}, which includes the address and the amount of the transfer. A transaction with n outputs has $0,\ldots,n-1$ output indices. Outputs from previous transactions are used as inputs in new transactions, which in turn create new outputs. In this way, transactions consume outputs and create new ones. For each output $o_y$ that is being used as the input of a transaction $t_y$, three data pieces are required:

\begin{itemize}
    \item Id of the transaction $t_x$ that created the output $o_y$ in an earlier time.
\item The index number of the output $o_y$ in the transaction $t_x$. 
\item Coin amount of the output.
\end{itemize}

Unlike many real-world transactions, which typically involve only one sender and one receiver, a Bitcoin transaction can involve multiple addresses. For example, a transaction can be one-to-one, one-to-many and many-to-one and many-to-many. A transaction with multiple inputs and multiple outputs is an interesting data type that we do not encounter in banking and finance; multiple addresses merge their coins and send them to multiple receiving addresses. 

 When a transaction has multiple input addresses, each address must sign the transaction separately using its associated private key. This ensures that the transaction is authorized by the owner of each input address, and prevents anyone from altering the transaction without the permission of the input owners.

The signatures on a transaction serve as proof that the transaction is legitimate and has been authorized by the owners of the input addresses. This is important because it allows the network to verify that the transaction is valid and should be included in the blockchain. Without these signatures, it would be possible for anyone to create and broadcast fake transactions, potentially leading to double spending or other types of fraud.

To receive Bitcoin, a user must create an address and share it with the sender. When a transaction includes multiple output addresses, the addresses do not need to be associated with the same person or entity. The same reasoning is less likely to apply to input addresses, because for a transaction to be created inputs addresses all must sign the the inputs with their private keys. In practice, most of the time, all input addresses belong to the same user. We can use this information to link addresses of the same user across transactions.

\begin{figure}
 \centering
\begin{subfigure}{0.45\linewidth}
  \centering
  \includegraphics[width=\linewidth]{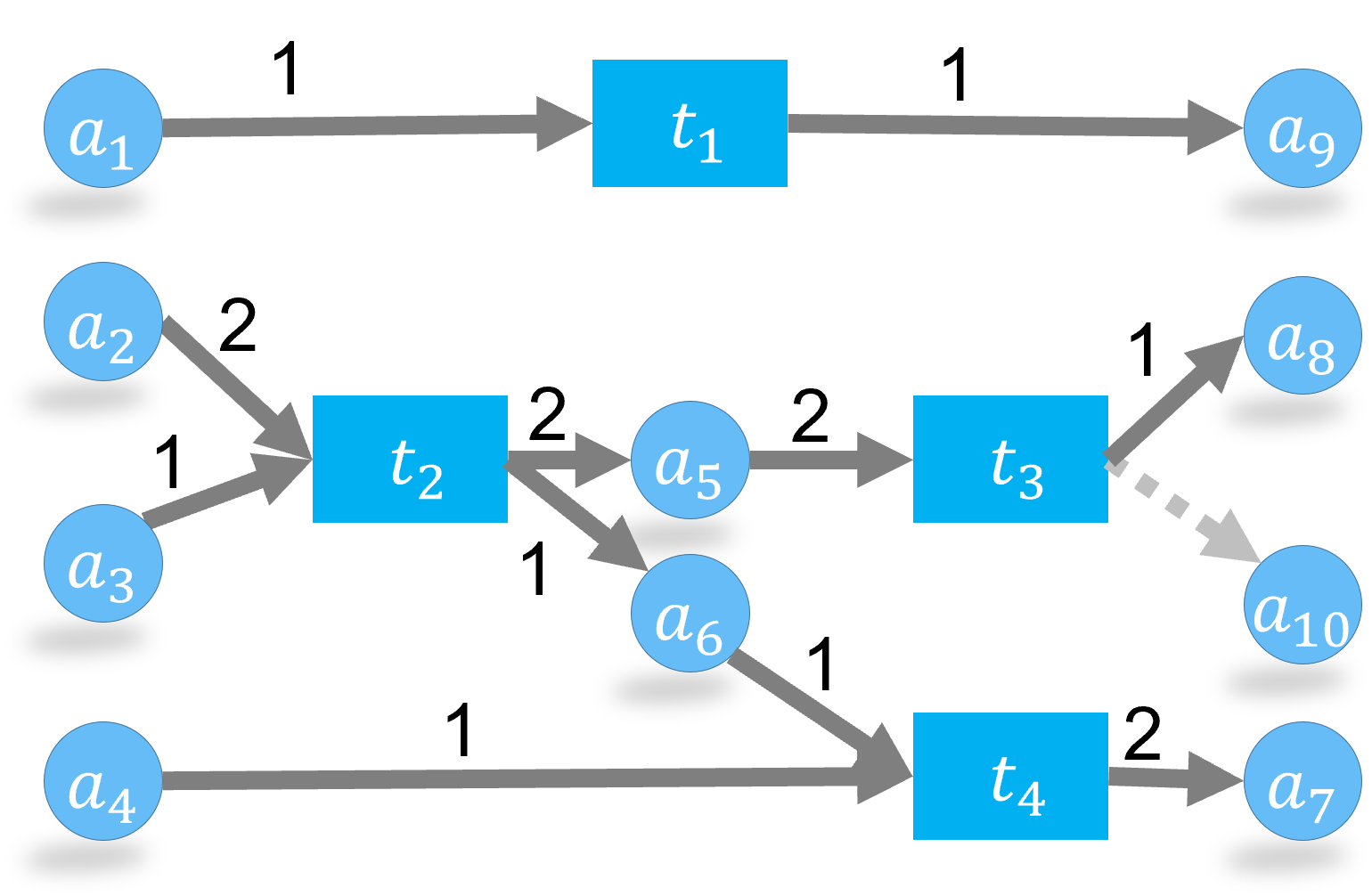}
  \caption{Four basic type of transactions involving 10 addresses.  Transaction $t_3$ sends funds to $a_{10}$ for transaction fees. Other transactions do not pay fees.}
  \label{fig:trans1}
\end{subfigure}%
~\text{ }~
\begin{subfigure}{0.45\linewidth}
  \centering
  \vspace{43px}
  \includegraphics[width=\linewidth]{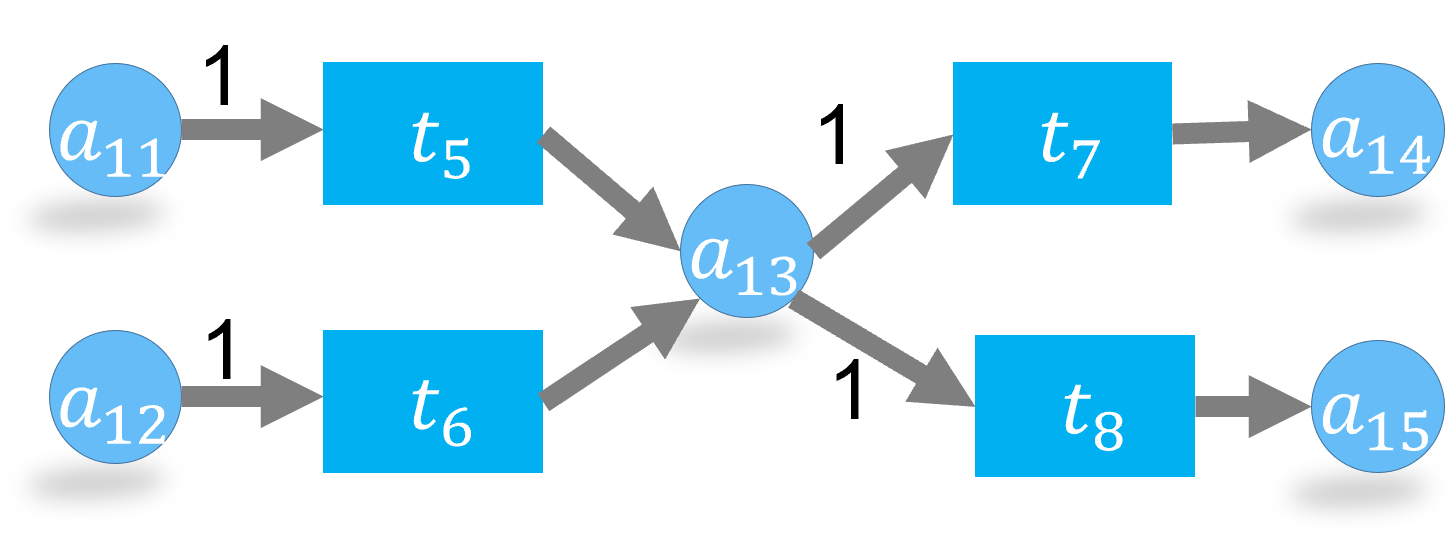}
  
  \caption{$a_{13}$ is spending bitcoins received from two transactions in two new transactions. An address can receive inputs from multiple transactions, but it has to spend each input completely when it spends.}
  \label{fig:trans2}
\end{subfigure}%
\caption{Bitcoin transactions and their spent and created outputs.}
\label{fig:transactions}
\end{figure}

Figure \ref{fig:trans1} shows four transactions where each transaction has a different shape. $t_1$ is a one-to-one transaction whereas $t_4$ is a many-to-many transaction. Some edges show coin amounts in Satoshi, which is the smallest unit of a Bitcoin, equal to one hundred millionth of a Bitcoin. 

In a transaction with multiple inputs/outputs, it is not possible to determine which input specifically funds a given output. This is similar to a lake where input rivers bring water and outgoing rivers (emissaries) take water out. It is not possible to trace a particular water drop back to its source river. This ambiguity was designed by Nakamoto. For example, in transaction $t_2$, it is not possible to know whether the amount in $a_6$ comes from $a_2$ or $a_3$. These multiple input/output transactions create interesting graphs.

We specifically want to emphasize two aspects of transactions. 

In Bitcoin, transaction fees are only implicitly specified by setting inputs and outputs. The difference between the total inputs and outputs is the transaction fee. While it is possible for a transaction to have no fee (i.e., the output amount is equal to the input amount), paying a fee can increase the likelihood that the transaction will be mined on the blockchain. Fees also help to prevent spam transactions on the network. In Figure~\ref{fig:transactions}, only transaction $t_3$ includes a fee, which we show as a dashed edge to the miner's address $a_{10}$.

Second, an address can receive transfers from multiple transactions and spend the outputs of these transactions separately.  For example, in Figure \ref{fig:trans2}, $a_{13}$ receives transfers from transactions $t_5$ and $t_6$. $a_{13}$ then spends two outputs (with $1+1$ coins in them) in transactions $t_7$ and $t_8$. This is possible because  $a_{13}$ had received multiple outputs. Now consider $a_5$ in Figure \ref{fig:trans1}. It also receives coins from transaction $t_4$, but it has to spend both coins at transaction $t_3$ because the two coins arrived in a single output. If $a_5$ only sends one coin, the miner will take the other as the transaction fee. That is, $a_{5}$ will lose any unspent amount. In practice, when a user has to spend a fraction of the received amount, it sends the remaining balance back to its address, or better, creates a new address, and sends it there. The new address is called a change address. This practice implies that the Bitcoin address of a user changes with each transaction. 

Manually creating transactions can be too tedious for ordinary users (See righto.com\footnote{\url{http://www.righto.com/2014/02/bitcoins-hard-way-using-raw-bitcoin.html}} for details). Online wallets are services provided by companies that allow users to easily send, receive, and exchange Bitcoin without having to deal with the technical details of transactions. However, these online wallets can be risky because they may leak the private keys of a user's account to malicious individuals, who can then steal their coins. Additionally, online wallets are subject to fraud, and users may not have the same level of control over their funds as they would if they managed their own wallet. \textquote{Not Your Keys, Not Your Crypto} is a common saying in the community.

The final step in creating a Bitcoin transaction is to broadcast the signed transaction, including all inputs and outputs, to the peer-to-peer network. This is done in the hopes that a miner will include the transaction in the next block added to the blockchain. Once the transaction is included in a block, it is considered confirmed and can no longer be reversed or altered. It is important to note that there is no guarantee that a miner will include a particular transaction in a block, and the inclusion of a transaction in the blockchain is not guaranteed.

\subsection{Verification and Confirmation}
\label{sec:miners}

Having covered addresses and transactions, we now turn our attention to how blockchain, particularly Bitcoin, uses them. A digital currency, such as Bitcoin, has two main problems: payment verification and payment confirmation.

\textbf{Payment verification} is the process of verifying that 1) the person making a payment has the necessary funds, and 2) the person intends to pay the specified amount. This is similar to verifying that someone wants to pay \$50 using authentic dollar bills. To verify a payment, a user presents their public key (to prove that the address belongs to them) and signs a signature with their private key (to confirm the amount). The transaction also includes as input a past transaction that created the output with the associated Bitcoin amount. This is called the proof of funds. For example, in Figure~\ref{fig:trans2}, $t_8$ lists $t_6$ as its input, showing the proof of funds.

These measures verify that a user has the balance and wants to make a payment. All network participants could have observed the transaction and recorded the new balances in an ideal peer-to-peer network. Otherwise, the spender can use the same bitcoins to make multiple payments. This issue is known as the {double spending} problem.

In reality, the network is faulty, and the news of a transaction may never reach some users. Furthermore, users could be malicious and lie about balances.  Due to these problems, distributed network users may never agree on who owns how many bitcoins and any payment would be a fraud risk. This issue is known as the Byzantine General's problem \cite{lamport1982byzantine}. As we mentioned earlier in Section \ref{sec:history}, early currencies used central authorities to solve this issue, but their fruitless efforts could only replace banks with companies. 

Nakamoto proposed a unique solution to the double-spending problem. Bitcoin uses a distributed public ledger to store time-stamped and linked blocks, which contain transactions. Each block carries a piece of information (i.e., block hash) about the previous block so that users can follow how the blockchain grows in time. Blocks have some constraints: as transferring big blocks among peers would clog the network, Bitcoin limits the block size to 1 MB. \footnote{In 2017 the {Segregated Witness} concept was activated to exclude transaction signatures from blocks so that miners could fit more transactions into a block.} When a new user joins the Bitcoin network, they must download the entire blockchain from other users (peers) on the network. This can take a significant amount of time, depending on the size of the blockchain. Once the download is complete, the user verifies all blocks and transactions, starting from the first block, in order to ensure that the blockchain is accurate and up-to-date. This process is necessary for the user to participate in the network and make transactions.

Block creation is limited to one per ten minutes (this is more a wish than a rule) through {Proof-of-Work}. Each block is computationally costly to create but easy to validate once created. Bitcoin Proof-of-Work requires finding a 32-bit number known as the \textit{nonce} through trial and error. In a sense, it is similar to buying a ticket to win a lottery. Using more computing power increases the probability of finding the number but does not guarantee it.  Users who work on finding confirmed blocks are called block {miners}.

\begin{figure}
\includegraphics{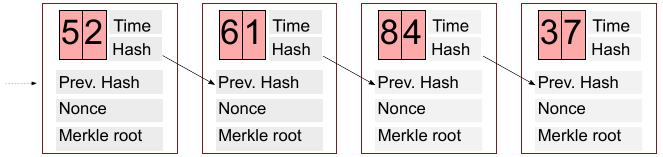}

\caption{A toy blockchain for the transactions given in Figure \ref{fig:transactions}. Each block is limited to have two transactions (shown with rectangles 1 to 8) and the coinbase transaction (not shown here).}
\label{fig:block}
\end{figure}

Proof-of-Work entails these steps: A miner receives a list of transactions from their peers, and chooses a selection of these transactions to include in the next block. Typically, transactions that pay a higher fee are more likely to be chosen. The miner then creates a binary tree (called a Merkle tree) from the selected transactions and computes the root hash of the tree. Several block-specific data pieces, such as the block's timestamp and the previous block's hash, are concatenated with the Merkle root hash to create a string. Note that adding the hash of the previous block creates a link between the current block and the previous one, and so on, forming a chain of blocks (see Figure~\ref{fig:block}). This is what gives the blockchain its name, as each new block is added to the end of the chain, forming a record of all previous blocks and transactions. This linking of blocks ensures the integrity of the blockchain, as any tampering with a previous block would result in a mismatch of hashes and would be immediately detected.

The miner then chooses a nonce value (an arbitrary integer such as 0) and appends it to the end of the string. The SHA-256 hash value of this string is compared against a target value. If the hash value is smaller than the target, the miner is said to have found a valid nonce and has successfully mined a block. If the SHA-256 hash value is greater than the target value, the miner has failed to find a valid nonce. In this case, the miner increments the nonce to 1 and repeats the process. The steps of creating the Merkle tree and concatenating the other data do not need to be repeated unless the transactions included in the block change. The miner continues to increment the nonce in the hope of eventually finding a valid nonce that satisfies the target value. As of 2022, the difficulty of mining a block is such that it typically takes around $10^{21}$ attempts to find a working nonce. This means that the process of mining a block is very difficult and requires a significant amount of computational power. As more miners join the network and the amount of computational power available increases, the difficulty of mining a block adjusts to maintain a consistent rate of block generation.

The difficulty \textit{target} is a globally determined value that is periodically adjusted depending on how long it took to find the last 2016 blocks. In theory, Bitcoin requires that a block take around 10 minutes, so 2016 blocks must take two weeks. However, depending on the number of miners and their computing resources, it may take more or less than the predicted two weeks.\footnote{See \url{ https://www.blockchain.com/charts} for the latest values.} The new difficulty value is increased or decreased accordingly to achieve 10 minutes per block. The difficulty tends to increase over time. However, several short-term decreases have occurred. 

In the POW scheme of Bitcoin, the miner spends a lot of effort to find the working nonce, but it takes only one hash computation and one comparison with the difficulty target to validate that the nonce works. In other words, this is quite an effective POW scheme.

Once a miner finds a block, they broadcast it to the rest of the network and hope that it will be accepted and added to the blockchain. As the information about the new block propagates throughout the network, users update their copies of the blockchain and set the new block as the latest one. When notified of a new block, miners should stop their ongoing mining efforts and update the difficulty target if necessary. They also remove any transactions that are included in the new block from their list of unconfirmed transactions, and update the \textquote{hash of the previous block} with the hash of the new block. Once these updates are completed, the miners can resume their mining efforts.

The first block of the chain, known as the genesis block, was mined by Nakamoto. Miner or not, a user in the system considers the longest chain the valid blockchain; it is also called the canonical chain.

In some cases, two miners may independently discover new blocks at roughly the same time. This results in a \textquote{fork} in the blockchain, where there are two competing versions of the chain that diverge from a common point. In this situation, users must choose which of the two forks to accept as the valid version of the blockchain. Typically, the fork that is perceived to require the most computational effort (i.e., the longest chain) is accepted as the main chain. Although longer forks are not common~\cite{lischke2016analyzing}, they can occur, as was the case in a well-known incident in 2013 where a version mistake in the Bitcoin protocol resulted in a fork that was resolved after 24 blocks.

Figure \ref{fig:block} shows how blocks can be created from the transactions that we gave in Figure \ref{fig:transactions}. Initially assume that the latest block is \shade{block 5-2} (i.e., \shade{Chain ..., 5-2}). Assume that after the \shade{block 5-2} is appended to the blockchain, \shade{block 6-1} and \shade{8-4} are mined at the same time. At this point, users and miners can choose any fork and continue their efforts because there are two forks of equal length and difficulty level.  

While miner \texttt{m} builds on \shade{Chain ..., 5-2, 6-1}, miner \texttt{n} searches for a new block on \shade{Chain ..., 5-2, 8-4}. After all, both chains are the same length, and choosing one over another is a kind of gambit.  Next, miner \texttt{m} also mines \shade{block 8-4}, so the fork becomes \shade{Chain ..., 5-2, 6-1, 8-4}. Note from Figure \ref{fig:transactions} that \shade{blocks 6-1} and \shade{8-4} contain different transactions, so they can be mined one after another. At this point, other miners in the system see that there are two forks, and \shade{Chain ..., 5-2, 6-1, 8-4} is the longest. They choose to build on this long chain. Otherwise, they risk their resources to be wasted on the shorter chain. Once the network miners start building on \shade{chain..., 5-2, 6-1, 8-4}, the chances of miner \texttt{n} mining a longer fork get even slimmer.  No rule stops miner \texttt{n} from continuing to work with its fork.  Blockchain assumes that this effort will be futile because the rest of the network will create a much longer chain than that of miner \texttt{n}.  

A transaction is tentatively considered \textit{confirmed} when it appears in a block. In practice, we deem a transaction to be secure after six confirmations, i.e., six blocks.

With Proof-of-Work, Bitcoin ensures that deliberately creating a fork to replace the main blockchain eventually becomes very expensive; a malicious user must mine new blocks faster than all other miners in the network combined. This means that the malicious miner must hold at least 51\% of all mining resources, which is not probable. However, mining power consolidation can complicate this simple equation. Researchers has shown that selfish miners can hide newly-generated blocks from the main blockchain to earn more bitcoins in block mining~\cite{eyal2014majority}. 

Although transactions are secured by the mechanisms we mentioned so far, malicious users have found multiple ways to scam wallets and users (See an excellent survey on all attacks by Saad et al.~\cite{saad2020exploring}). 

In transactions, we briefly mentioned that miners who mine blocks receive transaction fees. Transactions themselves do not have to pay fees, and miners can put non-paying transactions in a block as well. However, setting aside a fee increases the chances of a transaction being mined. As Bitcoin receives more transactions, waiting queues have become longer. For the future, a possible remedy is to increase the block size from 1MB to contain more transactions. Increasing the block size to reduce the fees has always been a contentious issue; after a period of much debate in 2017, Bitcoin Cash was created to use 8MB blocks. Bitcoin itself has, so far, resisted all calls to increase the block size.

In addition to transaction fees, the Bitcoin protocol provides a mining reward for each block that is successfully mined. Transaction fees plus the mining reward is the first transaction in a block, and it is known as the coinbase transaction. Mining reward is the only way that new bitcoins are created. When Nakamoto was the only miner on the network, the coinbase transaction was the only transaction included in each block. The mining reward started at 50 bitcoins per block and is halved every 210,000 blocks (four years). The total number of bitcoins that will be created through this process is limited to 21 million. After 2140 when mining reward becomes too small, transaction fees are expected to provide sufficient incentive for miners to continue securing the network.

\section{Changes and Improvements in Blockchain}
\label{sec:future}

As Bitcoin became popular, its limitations also became more visible. New generation blockchains learned from these limitations and improved over Bitcoin. Results range from minor tweaks to revolutionary ideas. In this section, we will go over the most important ones.

\subsection{Smart Contracts} 
Ethereum was created in 2015 to implement transactions and software code that execute conditions and rules. Those so-called smart contracts are written in proprietary coding languages, such as Solidity, and put to a network address by everyone to see and analyze. An analogy is the MYSQL snippets stored on a database. A contract clearly defines the functions that can be used and are guaranteed to work in a specified way. A simple example is an exchange service of assets $\mathcal{A}$ and $\mathcal{B}$. The order of transactions is as follows:
\begin{enumerate}
\item \label{step:2} Alice writes a contract with three functions: withdraw and exchange, and puts this to an address by including the contract code in an Ethereum transaction. Alice creates 100 $\mathcal{A}$ in the contract. The contract specifies that it will do an exchange for $1 ether = 5 \mathcal{A}$.  Alice signs the contract transaction with her private key.
\item Bob has ethers and wants to exchange them for $\mathcal{A}$s. Bob sends 20 ethers to the contract's address.
\item \label{step:3} The contract automatically accepts 20 ethers and assigns 100 $\mathcal{A}$s to Bob's address. This assignment is internal to the contract; a key-value pair of Bob-100 is added to the smart contract data storage.
\item The contract can accept sell/buy orders from other Ethereum addresses and act as a bridge between buyers and sellers.
\item Alice may use the withdraw function and receive 100 ethers from the contract.
\end{enumerate}

In this contract, Alice is selling an imaginary asset $\mathcal{A}$ that she created. The price specified in the contract reflects Alice's perception of the value of $\mathcal{A}$. Once the contract is created, it can be executed without any further involvement from Alice, ensuring that she cannot intervene, take Bob's ether, and disappear. In reality, contracts can be more complex and have been misused in the past. Formal verification of contracts to ensure that they function as intended is a growing area of research~\cite{bhargavan2016formal}.

Assets like the one described in this example are known as ERC20 and ERC721 tokens, which are named after the Ethereum Request for Comment (ERC) standards that they follow. ERC20 and ERC721 tokens are commonly used on the Ethereum blockchain to represent and transfer digital assets. ERC20 tokens are "fungible," meaning that they are interchangeable and each one has the same value as any other ERC20 token of the same type. In contrast, ERC721 tokens are "non-fungible," meaning that each one is unique and has its own distinct value. These tokens are similar to arcade tokens we used to buy at high school and insert into that helicopter game where you could shoot at ships directly or bounce bombs of the ground to do so. \footnote{\tiny Regretfully, the author could never remember the game's name.}

\begin{figure}
  \begin{center}
    \includegraphics[width=0.9\linewidth]{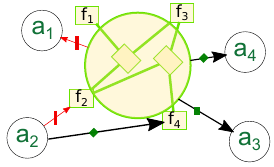}
  \end{center}
  \caption{Four addresses and a contract (it is an address as well). The contract has four functions and two decision boxes. Depending on the function used by $a_2$, and even the amount of transferred asset, $a_1$, or $a_4$ and $a_5$ can be linked to $a_2$. }
  \label{fig:contract}
\end{figure}

Smart contracts bring an exciting notion to graph analysis. A contract can be thought of as a template for future edges. Once a node creates an edge with a contract (e.g., Step \ref{step:2}),  the contract will create more edges (e.g., Step \ref{step:3}) that are predefined but constrained by conditions (if Bob sends less than 5 $\mathcal{B}$s, exchange will be refused). These edges are called internal transactions on Ethereum. Figure \ref{fig:contract} shows such a contract with its created edges. Edge templates such as this bring cascading behavior into mind. How can the graph be manipulated by planning to facilitate or restrict cascading behaviors? Such topics have been studied in Decentralized Finance, a promising field of financial activity on blockchains. Other than  Ethereum, smart contracts are expected to play a significant role in the Internet of Things \cite{christidis2016blockchains}. Once used by billions of connected devices, smart contracts will create large amounts of graphs data.

\subsection{Platform over Cryptocurrency} Bitcoin transactions transfer the digital asset of bitcoin currency. A transaction consists of addresses, amounts of bitcoin, and other data such as a timestamp. It is important to note that  bitcoin is just a financial application of blockchain technology.  The underlying structure of the blockchain can be used to represent and transfer any digital asset. This has led to the development of blockchain platforms that can be used to store and track legal documents, contracts, and even physical objects like diamonds. Blockchain platforms also have their own digital currencies to facilitate the mining of blocks, but transactions contain non-financial data and even software code, known as smart contracts. Various types of assets are transacted between addresses.

From a graph perspective, we use multilayer networks to represent asset networks on the blockchain. Graph nodes represent addresses of investors, and edges represent exchanges of various assets. This multilayer network contains valuable information about how different assets move, change hands, and affect each other's demand. For example, by 2021, the Ethereum blockchain had already developed a large ecosystem of smart contract-based tokens, decentralized exchanges, and other applications, which we cover next in Decentralized Finance.

\subsection{Decentralized Finance}
The growth of the multilayer network of assets on blockchain platforms has paved the way for the development of decentralized finance (DeFi), which refers to the use of blockchain technology to provide financial services in a decentralized manner, without the need for traditional intermediaries like banks. DeFi applications, which are built on top of blockchain networks like Ethereum, enable users to access a wide range of financial services, such as lending, borrowing, and trading, directly from their wallets. The rise of DeFi has facilitated the development of complex financial applications and services, such as stablecoins, decentralized exchanges, liquidity pools, and price oracles.

\textbf{Stablecoins} are a type of digital asset that is designed to maintain a stable value, typically by being pegged to a real-world asset like a fiat currency~\cite{moin2019classification}. A stablecoin is a token itself, but its price (e.g., 1USD per token) does not fluctuate. Some stablecoins use algorithmic mechanisms to maintain their value, while others are backed by assets such as gold or other cryptocurrencies. The failure of an algorithmic stablecoin called Terra in 2022~\footnote{See \url{https://www.chartalist.org/eth/StablecoinAnalysis.html} for the networks} put significant stress on the DeFi ecosystem and highlighted the need for ongoing research and development in this area. 

\textbf{Decentralized exchanges} (DEXs) are digital platforms that allow users to trade cryptocurrencies and other digital assets in a decentralized manner, without the need for a central authority or intermediary. DEXs are built on top of blockchain networks and use smart contracts to facilitate peer-to-peer transactions. This allows users to retain control over their assets and ensures that transactions are transparent and secure. DEXs often offer a wider range of trading pairs and lower fees than traditional centralized exchanges, making them an attractive option for many cryptocurrency traders and investors.

\textbf{Liquidity Pools} are pools of capital that are used to provide liquidity for certain markets or assets. In the context of decentralized finance (DeFi), liquidity pools are created by users who deposit their assets into a smart contract on a blockchain network. These assets are then used to facilitate trades on decentralized exchanges, allowing users to buy and sell assets more easily and efficiently. By providing liquidity to the market, liquidity pools help to reduce volatility and make it easier for users to access the assets they want to trade. Liquidity providers are typically compensated for their contributions to the pool, either through transaction fees or through a share of the assets in the pool.

\textbf{Oracles} are specialized software that provide external data to smart contracts on a blockchain network. Oracles are particularly useful for a stablecoin because it must learn the coin price in the real world to maintain its peg. However, stablecoin smart contracts, as all smart contracts, are limited in that they can only access data that is already stored on the blockchain. This means that they cannot access external data sources, such as real-world information about prices, weather, or other data that may be relevant to the contract. Blockchain oracles provide a way for smart contracts to access this external data and use it to execute the terms of the contract. This allows for the creation of more complex and versatile smart contracts that can interact with the real world in various ways.

DeFi continues to grow, with billions of dollars in assets under management. Researchers and developers are exploring the use of graph machine learning techniques to improve the robustness and security of DeFi applications, including early failure detection, influential node analysis, and other applications.

\subsection{Privacy Coins}

Privacy coins are cryptocurrencies that are specifically designed to increase the privacy of transactions on the blockchain. Examples of privacy coins include Dash, Monero and Zcash. Dash uses a protocol-level coin mixing mechanism to obscure the source of funds in transactions. Monero uses ring signatures to hide the sender and recipient addresses, as well as the amount of the transaction. Zcash uses zero-knowledge proofs to hide the sender and recipient addresses, but allows users to choose whether to use the shielded pool, where transaction details are hidden from the public, or the unshielded pool, where transactions are public as in Bitcoin. Despite a few early protocol mishaps, both Zcash and Monero are now quite unbreakable at hiding transaction data. 

Privacy coins make it difficult for outside observers to track the movement of funds on the blockchain, providing greater privacy and security for users.

\subsection{Decentralized Autonomous Organizations} 

\epigraph{When this happens, machines (hardware and software) become peers in the economy rather than mere tools. Imagine a world where drones that own themselves make deliveries, where autonomous software applications engage in virtual business such as buying and selling server time or even buying and selling stocks and bonds. One day, you may just be hired by a machine, as a one-time gig, or perhaps even for full-time employment. }{Adam Hayes \cite{hayesdao}}

Ethereum's smart contracts made it possible to create a system where actors get into contracts with each other to record matters, decisions, and results of decisions for everyone to see. The contracts encouraged the community to advocate for a future where Ethereum can be used to create an online democracy. This utopic future is discussed under the term \textbf{Decentralized Autonomous Organizations (DAO)}.   

The first attempt to create such a future was \textit{the DAO project}, which aimed to create an online investor fund. Joining members could vote on decisions to invest and earn money through their investments. Due to a hacking incident, the DAO became a traumatic experience that lost \$50M and led Ethereum to split into two. We will mention this story in Section \ref{sec:forkissues}. 

The DAO project left a bitter taste in the community and hindered the development of new DAOs. Although successful examples, such as the Maker/Dao stablecoin on Ethereum,  exist, there is not yet a consensus about what a DAO is and how it should use humans, robots, contracts, and incentives. Some argue that humans should not be involved in anything other than creating contracts, putting them online, and occasionally providing services to DAO (and getting paid). Others use a few humans as \textit{curators} that will manually filter incoming proposals to the DAO before a vote. We may add that even the term DAO is used in different meanings. However,  a consensus on DAO terms may be asking for too much.

Instead, a few influential developers have opined on what should be the common points of all DAOs. For example, the creator of Ethereum mentions \href{ https://blog.ethereum.org/2014/
05/06/daos-dacs-das-and-more-an-incomplete-terminology-guide/
}{{three points \cite{vitalikdao}}}. 
 
\begin{enumerate}
\item DAO is an entity that lives on the internet and exists autonomously.
\item DAO heavily relies on hiring individuals to perform certain tasks that the automaton itself cannot do.
\item DAO contains some internal property that is valuable in some way, and it can use that property as a mechanism for rewarding certain activities
\end{enumerate}

Opinions differ on how DAOs should function, but not on how influential they will be.  Consider this example from Hayes \cite{hayesdao}. A DAO acquires a car and puts its contract on Ethereum. The DAO investors can vote (without management) to send the car to work at the Dallas Fort Worth Airport. Riders pay the car on the blockchain, and the vehicle pays its investors dividends from what is left after gas, tax, maintenance, and insurance fees. This DAO can buy new cars, decide on their working sites and even connect cars to plan routes together. All of this is already possible through Ethereum.

\subsection{Fork Issues}
\label{sec:forkissues}
A short reading quickly reveals that most Blockchain technologies are forks of each other (See \href{http://mapofcoins.com/bitcoin}{{mapofcoins.com}} for a map). Blockchain uses \textbf{soft forks} to continue on the same main chain while changing a few aspects of the underlying technology. These are considered improvements or extensions. A soft fork is backward compatible and reflects community consensus on how the network should evolve.  \textbf{Hard forks}, on the other hand, creates a split in the main chain: two versions of the main chain are maintained by different groups of people. In a sense, it builds a new currency, technology, or community. The most famous hard fork happened in the Ethereum project in 2016 due to the DAO hacking. 

A hacker stole \$50M from \textit{the DAO project} which had raised \$150M from the community for a proof-of-concept investor fund. The DAO was hacked because of coding mistakes that allowed multiple refunds for the same investment. Hackers drained DAO while the community was watching the theft in real-time, helplessly.  Many developers wanted to roll back the transactions to forfeit the stolen amount. Disagreeing users rejected the rollback and stuck to the existing blockchain. The fork created two versions: Ethereum and Ethereum Classic. This fork started a discussion on how Ethereum should be governed. The \textbf{Code is law} proponents claim that any behavior that conforms to the Blockchain protocols should be accepted; a theft, once happened, cannot be punished by rolling back transactions. For legal aspects, see the article by \href{https://www.coindesk.com/code-is-law-not-quite-yet/}{{Abegg}} \cite{abegg}.

\subsection{Scalability Issues on Blockchain}

Bitcoin expects a block to be mined every 10 minutes, and it also imposes a block size of 1 MB, thereby limiting how many transactions Bitcoin can process in 24 hours. As a result, Bitcoin processes few transactions per second only. The payment method VISA, on the other hand, processes 2000 transactions per second, on average (See the \href{https://en.bitcoin.it/wiki/Scalability}{{bitcoin wiki}} for more details \cite{bitcoinscalability}.). 

Size and speed limitations have been eased in some other coins; Litecoin, for example, mines blocks every 2.5 minutes. The Bitcoin community has been discussing ways to increase the number of transactions. Bitcoin adopted a solution in 2017 with the Segregated Witness (SegWit) extension. Segregated Witness removes transaction signatures from the block, which reduces block size by 60\%. Further improvements would require increasing the block size itself, which the community still debates. See the \href{https://en.bitcoin.it/wiki/Block_size_limit_controversy}{{bitcoin wiki}} for details \cite{bitcoinsize}.

\subsection{First or Second Layer Technologies} Scalability efforts have culminated in second layer solutions, such as the Lightning Network~\cite{poon2016bitcoin}, where most of the transactions are executed off the blockchain. The first layer (i.e., the blockchain itself) only stores a summary of transactions that occur on the second layer.  Second layer networks, such as the Lightning Network, are a hot research area in network analysis that can be the topic of an extensive review article on its own. Due to space limitations, in this manuscript we will only teach blockchain networks that can be extracted from the first layer, i.e., the blockchain itself.

\subsection{Private Blockchains}
By definition, any user can join the Bitcoin blockchain, and all transactions are immutable and public. For corporate settings, this transparency means that rivals can learn company finances and buy/sale relationships. The \textbf{Hyperledger Project} was created to use blockchain in industrial settings. Supported by many organizations, Hyperledger offers membership services to choose blockchain participants and uses permissioned and even private blockchains. The project is hosted by the Linux Foundation and focuses on the blockchain's storage, capacity, and availability aspects. Hyperledger users are allowed to choose their consensus and mining approaches. For more on Hyperledger, see \cite{aggarwal2021hyperledger}.

 \subsection{Proof-of-X} 
 
 A major criticism of Bitcoin is that while miners are racing to find the next block, no thought goes into how much power and resource is wasted. There can be thousands of other miners whose computations are wasted for a single winner at every block. In 2014, it was estimated that one bitcoin cost 15.9 gallons of gasoline. \footnote{\url{http://www.coindesk.com/carbon-footprint-bitcoin/}} It will only get worse as more miners join the race. As Swanson notes, Bitcoin is, in reality, a \textquote{peer-to-peer heat engine}~\cite{swanson2014learning}. However, as Eric Jennings writes \textquote{the cost for having no central authority is the cost of that wasted energy}. 

Alternatives to POW aim to reduce computational inefficiency by relaxing a few assumptions made by Bitcoin. 
Proof-of-X is an umbrella term that covers POW alternatives in block mining. Each alternative expects miners to prove that they have done enough work or spent enough wealth before creating the block. In POW, the work is the mining computations, and the proof is the hash value.
Instead of doing some work, some wealth can be destroyed (as a kind of sacrifice) to mine a block. \textit{Proof-of-Stake} considers coin-age as wealth; 10 2-year-old coins create a wealth of $10 \times 2 =20$ stakes. The block whose miner has the highest stake becomes the next block in the chain. Once coins are used as a sacrifice, their age becomes zero. Coins will have to accumulate their wealth again. \textit{Proof-of-Burn} goes a step further and indeed sacrifices coins. In the scheme, a miner first creates a transaction and sends some coins to a \textquote{verifiably unspendable} address. These coins are called burned, but other than the sender, no one in the network is yet aware that the receiving address is an invalid/unusable one. Afterward, the miner creates a block and shows the proof of burnt coins. The proof is a script that shows how the address was created through erroneous computations. The miner who has burnt the highest number of coins can mine the next block. In this scheme, burning coins to collect transaction fees is only viable if transaction fees/rewards are high enough. Furthermore, the total coin supply will decrease in time. In \textit{Proof-of-Disk} nodes that commit computer memory to maintain the network and perform network functions are chosen as miners.

There is an essential shortcoming in POW alternatives. Maliciousness does not have repercussions. If miners decide to game the system by supporting every competing fork, Proof-of-Work becomes too expensive due to the required computations. Other schemes are not as effective against miner malice.

Consider the case with two competing forks in the blockchain. In Stake or Burn-based proofs, miners can create multiple blocks built on different forks using the same (burned or aged) coins. Eventually, one of the forks will be the longest and become the main chain. Regardless of which one is chosen, the miner's block will have taken its place in the chain, and the miner will receive block rewards. This is known as the \textit{nothing as stake problem}; the miner does not lose anything by gaming the system.

 \section{Blockchain Network Analysis}
\label{sec:graphs}
We can categorize blockchains into two classes in terms of data models in their transaction networks: \textbf{account based} (e.g., Ethereum), \textbf{unspent transaction output (UTXO)  based} (e.g., Bitcoin, Litecoin) blockchains, and DAG (directed acyclic graph) based blockchains.

Traditionally cryptocurrencies are UTXO based, whereas platforms are account-based. The difference between UTXO and account-based models has a profound impact on blockchain networks. 

In account-based blockchains, an account (i.e., address) can spend a fraction of its coins and keep the remaining balance. An analogy to account-based blockchains is a bank account that makes payments and keeps the remaining balance in the account. In these blockchains, a transaction has exactly one input and one output address. We may use an address to receive and send coins multiple times. The resulting network is similar to traditional social networks, implying that we can directly apply social network analysis tools to account networks.

On UTXO based blockchains, nodes (i.e., addresses) are dynamic, which complicates network analysis. As Bitcoin became popular, research works analyzed the network for coin price predictions. Various features, such as mean account balance, the number of new edges, and clustering coefficients, have been used in research works  \cite{sorgente2014reaction,greaves2015using}. Other than features, network flows \cite{yang2015bitcoin} and temporal behavior of the network \cite{kondor2014inferring} have also been used in predictions. 

 \begin{table}[!ht]
 \caption{Blockchain types. The private/public columns indicate block mining permissions. IOTA and Ripple are maintained by consortiums that own the mining rights. Many projects (such as Ripple and IOTA) developed smart contract functionality many years after their launch. Although IOTA uses a directed acyclic graph instead of a canonical blockchain, the IOTA transaction model is UTXO based. Privacy coins Monero and Zcash use UTXO models with cryptographic security.}
    \label{tab:blockchainTypes}
    \centering
    \begin{tabular}{c c c c c c c c}
      
         &UTXO&Account&DAG&Platform&Cryptocurrency&Private&Public \\
       \midrule
         Bitcoin&\checkmark&&&&\checkmark&&\checkmark\\
        \rowcolor{Gray} ZCash& \checkmark&&&&\checkmark&&\checkmark\\
         Monero&\checkmark &&&&\checkmark&&\checkmark\\
         \rowcolor{Gray}  Ripple& &\checkmark&&\checkmark&&\checkmark&\\
           Ethereum&&\checkmark&&\checkmark&&&\checkmark\\
        \rowcolor{Gray} IOTA& \checkmark&&\checkmark&\checkmark&&\checkmark&\\
         \bottomrule
    \end{tabular}
\end{table}

Table~\ref{tab:blockchainTypes} shows a list of blockchains and their types. 2017 version of this primer had only discussed Bitcoin networks. In 2022, Bitcoin is still the main blockchain of discussion, and this primer again focuses on Bitcoin. However, influential blockchains that do not use the UTXO model have appeared. For in-depth information on UTXO, account and DAG based blockchains, we refer the reader to our recent public manuscript~\cite{gurcan2021blockchain}.

Studies in network features show that since 2010 the Bitcoin network can be considered a scale-free network \cite{lischke2016analyzing}. In and out-degree distributions of the transactions graphs are highly heterogeneous and show disassortative behavior \cite{kondor2014rich}. Active entities on the network change frequently, but there are consistently active entities \cite{ober2013structure}. The most central nodes in the network are coin exchange sites \cite{baumann2014exploring}.

\subsection*{UTXO Graph Model}
\label{sec:contentgraphs}
UTXO (e.g., Bitcoin, Litecoin, Monero, Zcash) transaction networks are modeled in three forms: transaction, address, and entity/user graphs. 

\begin{wrapfigure}{r}{0.4\textwidth}
  \begin{center}
    \includegraphics[width=\linewidth]{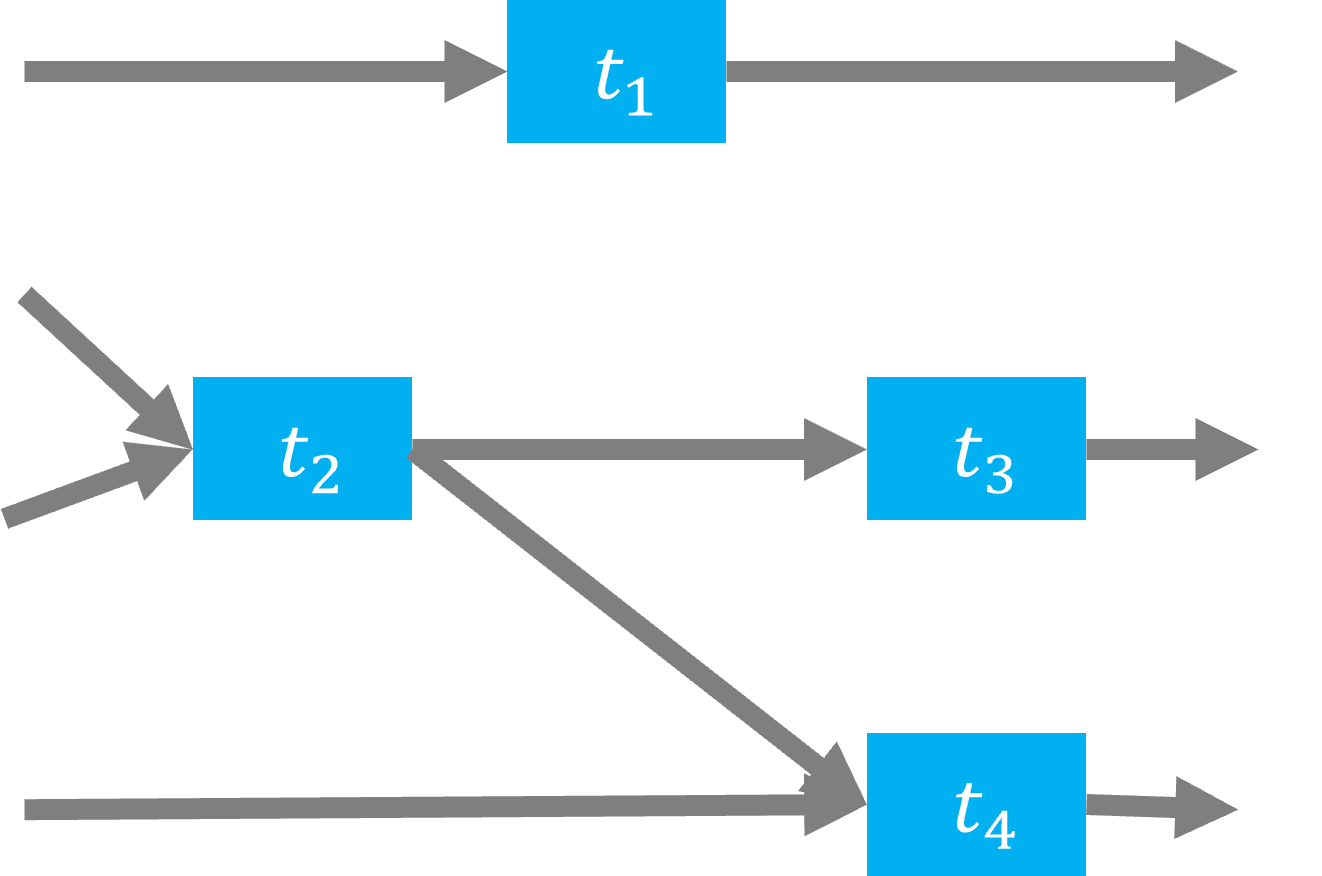}
  \end{center}
  \caption{A graph of the 8 transactions from Figure \ref{fig:trans1}. Transactions are ordered on the horizontal axis by their appearance in blocks (see Figure \ref{fig:block}).}
  \label{fig:graphtr}
\end{wrapfigure}

 Transaction graphs omit address nodes from the transaction network and create edges among transactions only.  Figure~\ref{fig:graphtr} shows the transaction graph of the network shown in Figure~\ref{fig:transactions}. The most important aspect of the transaction graph is that a node (a transaction), by definition, can appear only once. There will be no future edges that reuse a transaction node.

The \textbf{transaction graph} contains far fewer nodes than the network it models. We can immediately observe a few drawbacks from Figure~\ref{fig:graphtr}. Unspent transaction outputs are not visible; we cannot know how many outputs are there in $t_1$, $t_3$  and $t_4$. In Bitcoin, many outputs stay unspent for years; the transaction graph will ignore all of them. Second, the transaction graph cannot model address reuse; when a past address receives coins again from a transaction, we cannot create the edge because that would introduce a cycle in the graph.

The advantages of the transaction graph are multiple. First, we may be more interested in analyzing transactions than addresses. For example, anti-money laundering tools aim at detecting mixing transactions, and once they are found, we can analyze the involved addresses next. Many chain analysis companies focus their efforts on identifying transactions that are used in e-crime. Second, the graph order (node count) and size (edge count), useful for large-scale network analysis, are reduced from the blockchain network. In UTXO networks, transaction nodes are typically less than half the number of address nodes. For example, Bitcoin contains 400K-800K unique daily addresses but 200K-400K transactions only. However, the real advantage of the transaction graph is its reduced size. As we will explain in the next section, the address graph contains many more edges than the transaction graph. 

\begin{figure*}[h!]
\captionsetup[subfigure]{position=b}
\centering
\subcaptionbox{Address graph creates edges between all input-output address pairs. The resulting graph may have millions of edges due to a single transaction. \label{fig:graphadd}}{\includegraphics[width=.45\linewidth]{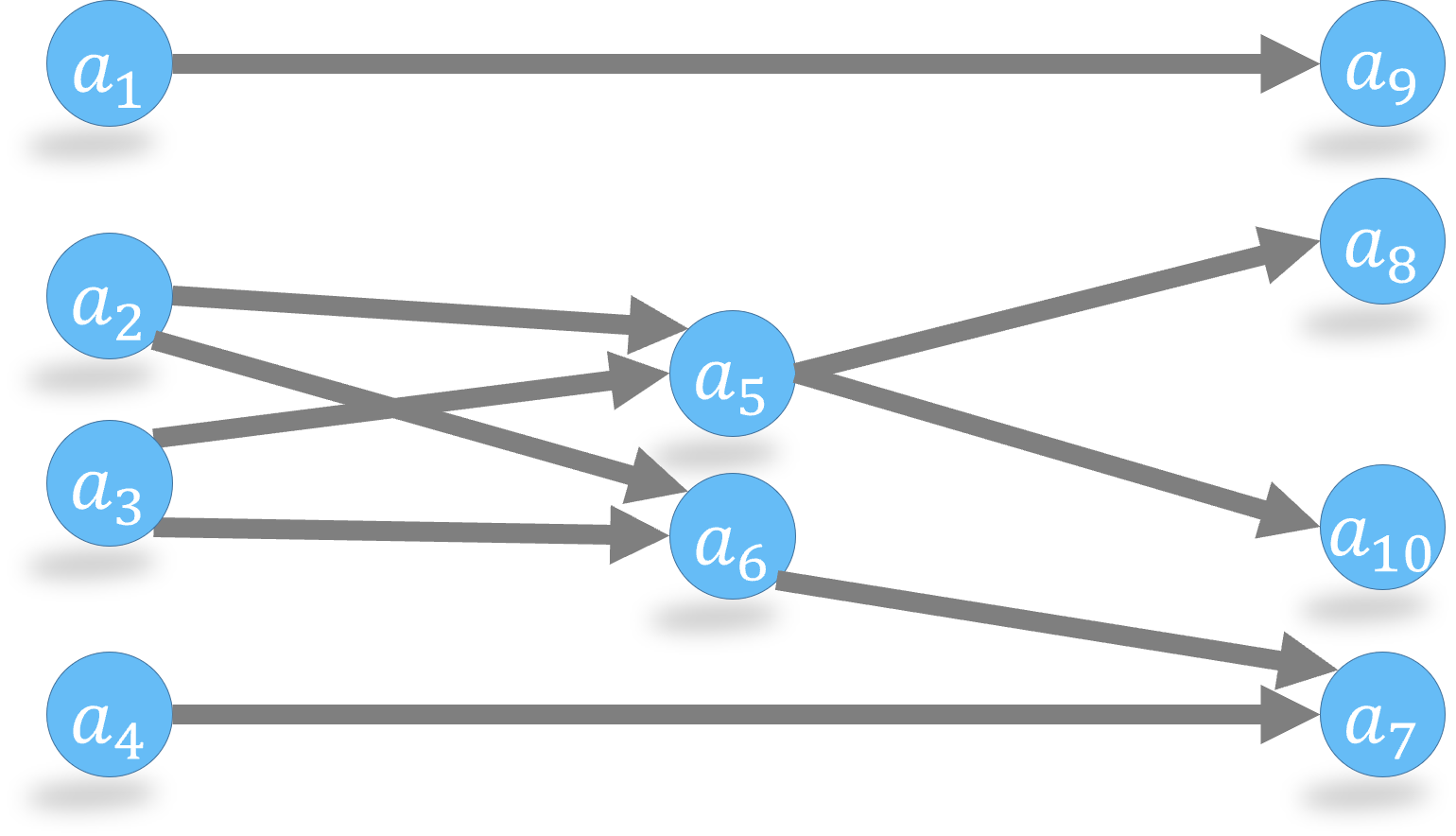}}
\text{ }
\subcaptionbox{Entity graph with owners of addresses shown as super nodes. Nodes in a transaction graph cannot create new edges in the future.\label{fig:graphent}}{\includegraphics[width=.45\linewidth]{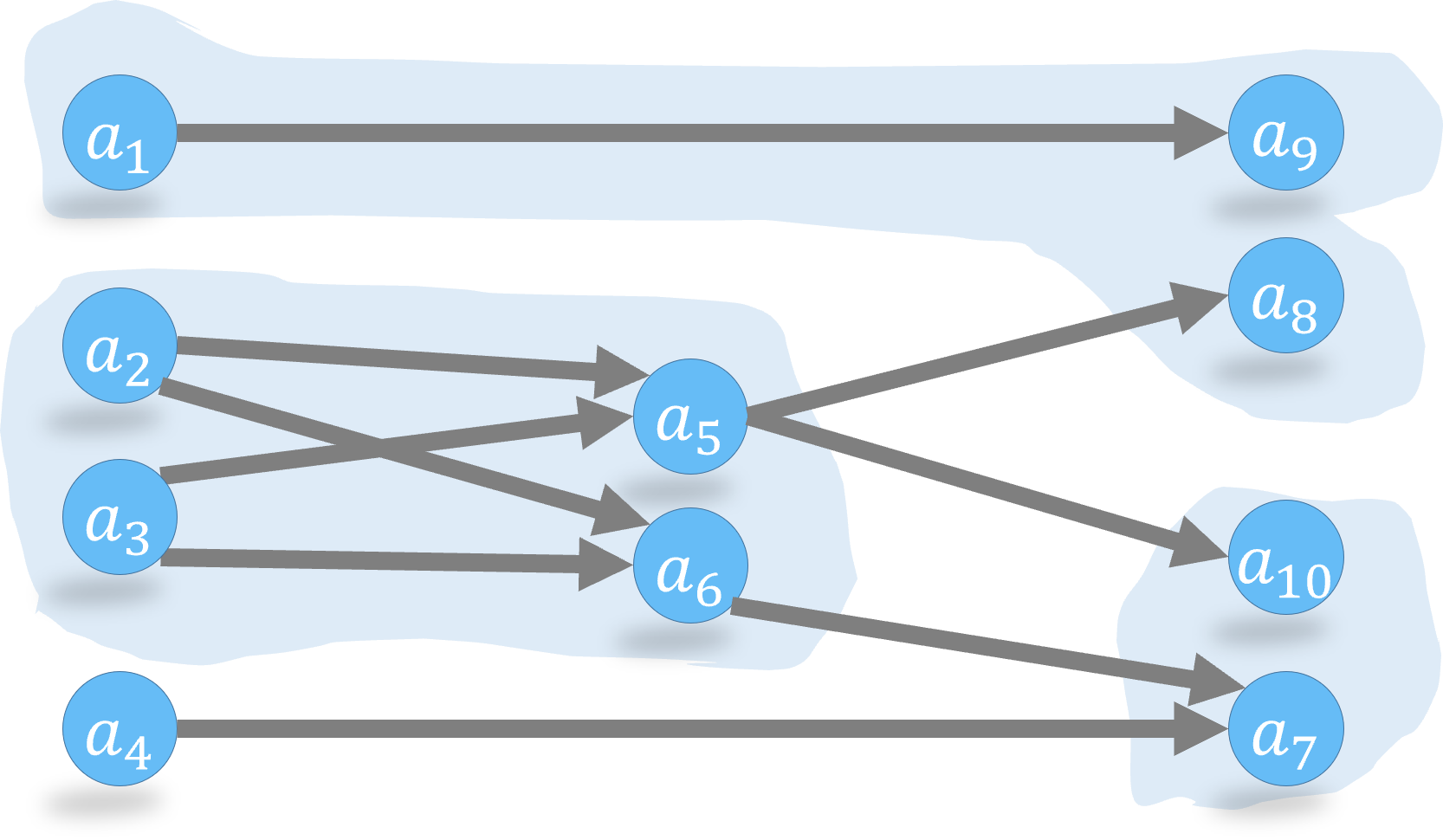}} 
\caption{Blockchain address and entity graphs for the transactions of Figure~\ref{fig:transactions}}
\label{fig:graphs}
\end{figure*}

The \textbf{address graph} (Figure~\ref{fig:graphadd}) is the most commonly used graph model for UTXO networks. The address graph omits transactions and creates edges between addresses only. Address nodes may appear multiple times, which implies that addresses may create new transactions or receive coins from new transactions in the future. Figure~\ref{fig:graphadd} shows the address graph of our recurring example from Figure~\ref{fig:transactions}. As in the transaction graph, each edge carries a weight (i.e., bitcoin amount), but multiple edges can exist between addresses.

Furthermore, if the change is sent back to the input address, the graph contains self-loops. In practice, however, an address is used to send assets only once. As a community rule, {address reuse}  is generally avoided to prevent transaction privacy attacks. Due to this, loops and multi edges are rare in the address graph.

Address graphs are larger than transaction graphs in node and edge counts. A UTXO transaction does not explicitly create an edge between input and output addresses in the blockchain transaction network. When omitting the intermediate transaction node, we cannot know how to connect input-output address pairs. As a result, we must create an edge between every pair. For example, as shown in edges from $a_2$ and $a_3$ to nodes $a_5$ and $a_6$, all input addresses contained within a transaction are deemed to have edges to each output address. If there are few addresses in the transaction, this may not be a big problem. However, large transactions can easily end up creating millions of edges. For example, the highest number of inputs in a Bitcoin transaction was 20000 (821 on Litecoin), whereas the highest number of outputs in Bitcoin was 13107 (5094 on Litecoin). The address graph approach will have to create $20000\times 13107$ edges for the largest Bitcoin transaction.

Graph size is not the only problem. The address graph loses the association of input or output addresses. For example, the address graph in Figure~\ref{fig:graphadd} loses the information that edges $a_2$ and $a_3$ were used in a single transaction; address graph edges would be identical if the addresses had used two separate transactions to transfer coins to $a_5$ and $a_6$. This issue can be solved by adding an attribute (e.g., transaction id) to the edge. 

Differing from others, the \textbf{entity graph} (Figure~\ref{fig:graphent}) tries to find which addresses are owned by the same entity. These efforts are also known as user/entity clustering.  Some entities, such as Wikileaks, publicly advertise its address. Some other addresses can be found on mailing lists, forums, etc. publicizing a user's address might provide benefits; if the user is known (i.e., trusted), transactions from her address can be accepted by vendors without waiting for six blocks. These known and trusted addresses are called {green/marked addresses}. 

In general, there is not a clear-cut method to link addresses of a user together. Some works have proposed graph algorithms to cluster users \cite{ron2013quantitative}. Despite having false positives,  heuristics have been widely used by the community and in research works. These heuristics are peeling chain \cite{meiklejohn2013fistful}, {change closure} \cite{androulaki2013evaluating}, {idioms of use},  {transitive closure} and Ip clustering  \cite{ortega2013bitcoin}.

The \textbf{peeling chain} is a technique often used by criminals to divide and hide the origins of funds on the blockchain. In a peeling chain, an initial address containing a certain amount of bitcoins is used. A small portion of these funds is then peeled off and sent to a new address, called a change address. This process is repeated multiple times, with the remaining funds from each transaction being peeled off and sent to a new change address. This continues until the initial funds have been divided into many small amounts and transferred to a large number of addresses, all of which are expected to be controlled by the same entity.

\textbf{Change closure} follows the community practice of sending whatever change remains from a transaction to a new address owned by the spender. For this heuristic, the change address should never appear before and never be reused to receive payments. In Figure \ref{fig:graphadd} after paying $a_{10}$, $a_5$ sends the change to $a_8$. Both $a_5$ and $a_8$ may belong to the same entity.

\textbf{Idioms of use} posits that all input addresses in a transaction should belong to the same entity because only the owner could have signed the inputs with associated private keys. By this heuristic, $a_2$ and $a_3$ from Figure \ref{fig:transactions} belong to the same user. 

\textbf{Transitive closure} extends Idioms of Use: if a transaction has inputs from $a_x$ and $a_y$, whereas another transaction has from  $a_x$ and $a_z$, $a_y$ and $a_z$ belong to the same entity.

\textbf{IP clustering }is a technique that uses network-level information to identify clusters of entities on the blockchain~\cite{biryukov2019deanonymization}. In the context of Bitcoin, this technique involves identifying addresses that use the same IP address and grouping them together into clusters. This can be useful for identifying the entities behind a group of addresses, as it is likely that addresses belonging to the same cluster are controlled by the same person or entity.

IP clustering has been proposed as a way to deanonymize Bitcoin addresses and improve the ability of law enforcement agencies and regulators to track the movement of funds on the blockchain. However, this technique is not foolproof, as users can easily use VPNs or other methods to mask their IP addresses and prevent clustering. Additionally, the use of clustering to identify entities on the blockchain raises privacy concerns, as it potentially allows outsiders to link addresses to real-world identities. As a result, the use of IP clustering for deanonymization purposes is an active area of research and debate.

By nature, all user clustering heuristics are error-prone. Some community practices further complicate the issue. For example, online wallets such as coinbase.com buy/sell coins among their customers without using transactions; ownership of an address is changed by transferring the associated private keys to another user. Although the user associated with the address changes, nothing gets recorded in the blockchain. 

A measure to prevent matching addresses to users is known as \textbf{CoinJoin}  \cite{maxwell2013coinjoin}. The key idea is to use a central server that mixes inputs from multiple users. Only the server knows the mapping between inputs and outputs.  Mistakenly, the Idioms of Use heuristic would mark all these input addresses to belong to the same user. Relying on a central server is a major weakness in CoinJoin, but there are attempts to develop viable alternatives \cite{ruffing2014coinshuffle}. Specifically, privacy coins ZCash, Monero, and Dash have seen increasing adoption since 2019. In theory, Monero, Dash, and ZCash privacy coins are  UTXO blockchains; their Blockchain networks are similar to Bitcoin's. In practice, privacy coins employ cryptographic techniques to hide node and edge attributes in the blockchain network. For example, ZCash hides information in its shielded pool, whereas Monero adds decoy UTXOs to the input UTXO set.  

The increasing usage of coins in crime has necessitated finding users/entities behind addresses. For example, ransom software that encrypts hard drives uses Bitcoin as a medium for ransom payments. Coins have also been stolen (i.e., transferred to addresses of thieves) from online wallets such as Mt.Gox. Companies such as Elliptic and  Numisight develop Blockchain graph analysis tools to track these crime-related coins. In research works, BitIodine \cite{spagnuolo2014bitiodine} and McGinn \cite{mcginn2016visualizing} offer visualization tools. GraphSense \cite{haslhofer2016bitcoin} is an analytics platform that also provides path search on transaction graphs. Chartalist (\url{https://github.com/cakcora/chartalist}) provides labeled graph datasets for graph machine learning on blockchains.

With so many works on identifying entities behind addresses, privacy research on Blockchains has also become an exciting topic. See \cite{lai2021survey} for a recent study on user privacy on Bitcoin.

 \section{Conclusion}
 We have presented a holistic view on blockchains and described aspects that affect graph mining on blockchain transaction networks. In this version of the manuscript, we have updated address and transaction details of blockchains and discussed the emergence of blockchain platforms with their asset networks. 

 \section{Next Steps}
 The following (links in the PDF) resources will be useful for your next steps.
\begin{itemize}
    \item Survey: Blockchain networks: Data structures of Bitcoin, Monero, Zcash, Ethereum, Ripple, and Iota \newline \url{https://wires.onlinelibrary.wiley.com/share/WKHPKR6MYBWT9UYVKRJB?target=10.1002/widm.1436}.
    \item Tutorial: WWW Conference,  {Cryptoasset Analytics: Data, Fundamental Concepts, and Open Source Tools} \newline \url{https://www.youtube.com/watch?v=hMV9zSQOO98}.
    \item Course: {Udemy, Data Science on Blockchains}  \newline \url{https://www.udemy.com/course/data-science-on-blockchains/learn/}.
    \item Data: {Chartalist Graph Machine Learning Repository}  \newline \url{https://github.com/cakcora/chartalist}.
    
\end{itemize}
 
\bibliographystyle{ACM-Reference-Format}
\bibliography{main}
\end{document}